# Energy Aware Wireless System based Software Defined Radio

H. Salman[1,2], R. Balatiah[1], A. Masri[1] and YAS. Dama[1]
[1]Department of Telecommunication and Electrical Engineering, An-Najah National University, Nablus, Palestine
[2]Department of Electrical and Electronics Engineering, Istanbul Medipol University, Istanbul, Turkey
Email: hmysalman@st.medipol.edu.tr, ahmed.masri@najah.edu, yasdama@najah.edu

*Abstract*— Development of green telecommunication systems is already being considered highly attractive by standard bodies and recently is attracting research attention. While most of the research focuses on modeling and simulation, in this work we implement a lab setup to test an energy aware wireless system based on software defined radio and solar energy power system. In addition, we proposed an energy aware adaptive modulation algorithm that considers the state of charge of the solar energy batteries before setting up the modulation order. Moreover, the algorithm adapts to user preferences between the connectivity mode and the quality mode.

## I. INTRODUCTION

Future mobile and wireless communications envision magnitudes of increase in data rates, bandwidth, coverage and connectivity, with a massive reduction in round trip latency and with great focus on reduction in energy consumption [1]. A sustainable wireless network should not only be Spectrally Efficient (SE) but also Energy Efficient (EE) [2]. For that, development of Green Telecom (GT) is already being considered by standard bodies, where reduction in energy usage will be by almost 90% [3].

In fact, Information and Communication Technologies (ICT) hold the responsibility for a significant proportion of global energy consumption. In ICT systems core and access network are estimated to be equal in power consumption by 2017 [4][5]. In 2009 Deutsche Telekom forecasted a 12-fold increase in the power consumption by network core within a decade [4]. Power Consumption by backbone networks (routers, fibers, transmission) is approximately 12% of electricity consumption in broadband enabled countries and by 2020, it is estimated to increase to about 20% [5].

Extensive research work has been already carried out all around the world in the domain of GT [3][4][5][6] and [7]. In [7], authors emphasized on how GT networks are attracting an increasing attention by researchers in GT domain. This is due to their critical role in minimizing the overall consumption of energy by applying EE techniques and using renewable energy resources, as well as, eco-friendly consumables and wireless energy harvesting techniques.

There are global examples of using solar systems for electrification of telecom systems. Indonesian operator PT Telecmunikasi cellular (Telkomcel) is using latest generation low power consumption radio blocks which are powered by solar technology from Ericsson to provide macro coverage in Sumatra and rural areas of Indonesia [8]. Another example in [9], Jordan Telecom implements solar energy project for telecom tower in Karak area which is a hybrid system that contains solar power panels, wind turbine and diesel engine generator. Moreover, they are installing solar systems for outdoor sites, where there is no need for a/c unites and the average load is 300 W to 1400 W including the consumption of fans.

Authors in [10] designed energy-scalable SDRs through introducing an energy aware cross-layer radio management framework. However, their work is limited to design and simulation without real implementation. Moreover, they do not consider the usage of Renewable Energy Resources (RER). Where, using RER for powering up wireless systems makes the continuous functionality/connectivity of such systems vulnerable to several factors, mainly related to weather conditions. As these conditions are changing continuously and sometimes unexpectedly, the amount of energy available for powering up wireless systems will be also changing and sometimes unpredictable. Based on this well-known fact, conventional wireless systems-based RER will be susceptible to communication interruption when the available energy is no more enough to satisfy the system requirements.

A tradeoff between SE and EE is proposed and a general metric SEE (Spectral/Energy Efficiency) which quantifies the preference of SE or EE, has been formulated in [11]. While in [12] the energy consumption of M-PSK modulation schemes is investigated under various fading channels. More explicitly, the total energy required to transmit each bit of information is formulated into an objective function. In both works [11] [12], the wireless systems are assumed to be powered up using conventional sources of power. on the contrary in this paper, we setup a lab test for Energy aware Wireless System based Software Defined Radio (EWS-SDR). Trade-off between Connectivity Efficient mode (CE) and Quality Efficient mode (QE) based algorithm is proposed and tested. In specific, an adaptive modulation based available energy technique is used.

The rest of paper is organized as follow: In section II, we describe our system model and hardware lab setup. Then, in section III, the USRP N200 and solar power calculations are detailed. Next, our EWS-SDR adaptive algorithm is proposed and detailed on section IV. Then, Lab tests and results are discussed in section V. Finally, in section VI we present our conclusions.

## II. SYSTEM MODEL

Our system consists of a pair of wireless radio devices that are assumed to be powered up using solar energy system. Each radio device is based on software defined radio system and is connected through Gigabit Ethernet switch to a computer with MATLAB-Simulink software for signal processing and performance evaluation. Figure 1 presents our lab setup.

Control information is exchanged between the transmitter and receiver through a feedback channel working on an out-of-band frequency. The feedback channel is implemented using BPSK modulation technique to guarantee connectivity [12]. The authors of [12] formulated an objective function (energy per bit) which is fundamentally dependent on modulation order (M) and Signal to Noise Ratio (SNR) which is determined by the fading conditions at a given target Bit Error Rate (BER). Based on their results, Binary Phase Shift Keying (BPSK) is the best choice for transmission range longer than 20 m and is usually preferred in order to achieve the minimum energy consumption per bit.

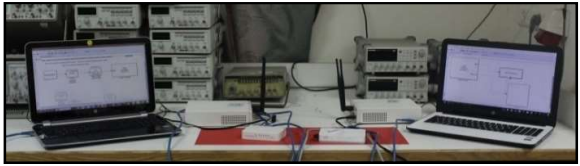

Fig. **1**. EWS-SDR setup

### A. Hardware Components

In our lab test we use a pair of Universal Software Radio Peripherals (USRP) Networking series (N200) to represent our wireless transceivers working in the Industrial, scientific and Medical (ISM) radio frequency band. Moreover, each USRP N200 is equipped with two dipole antennas, one for data and the other for feedback control channel. Table I, lists the USRP N200 specifications [13].

In section III, we detail the description about the required solar system for powering up our USRP devices.

TABLE I.  USRP N200 SPECIFICATIONS

| USRP-N200 | Specifications | | |
|---|---|---|---|
| | item | Value | Unit |
| Power | DC input | 6 | V |
| | Current Consumption | 1.3 | A |
| Conversion performance and clock | ADC sample rate | 100 | MS/s |
| | ADC resolution | 14 | Bits |
| | DAC sample rate | 400 | MS/s |
| | DAC resolution | 16 | Bits |
| | Frequency Accuracy | 2.5 | ppm |

| USRP-N200 | Specifications | | |
|---|---|---|---|
| | item | Value | Unit |
| Physical | Operating temperature | 0-55 | C |
| | Dimensions (L*W*H) | 22*16*5 | cm |
| | Weight | 1.2 | Kg |
| RF Daughterboard XCVR 2450 | Frequency | 2.4, 5 | GHz |
| | Power Output | 15 | dBm |
| | Current Consumption | 2.3 | A |
| | Receiver noise figure | 5 | dB |
| Antenna VERT2450 | Profile | 197 | mm |
| | Operating frequency | 2.4~2.5/5.15~5.35/ 5.725~5.85 | GHz |
| | Type | Dipole | |
| | Polarization | Linear | |
| | Radiation | Toroidal | |
| | Peak Gain | 2.0 | dBi |
| | Impedance | 50 | ohm |
| | VSWR | 2.0:1 Max | |
| | Swivel | 90 | degree |
| | Connector | SMA | (M) |

### B. Software Components

In the transmitter side, the message is generated then modulated using M-PSK modulator. After that, a square root raised cosine is used to shape our symbols. Finally, the filtered symbols transmitted over the air using the SDRu transmitter system object that connects our software part to the USRP hardware as shown in Fig 2.

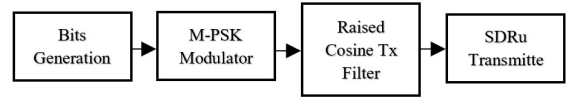

Fig. **2**. The Simulink top-level structure of M-PSK transmitter with USRP N200 Hardware

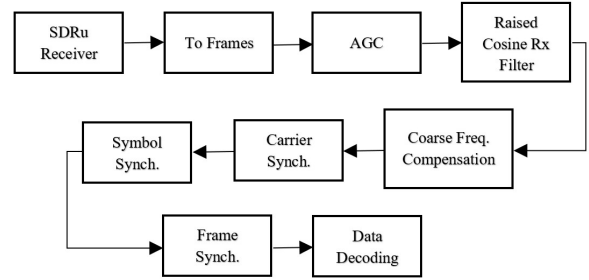

Fig. **3**. The detailed structures of the M-PSK Receiver subsystem.

Figure 3 shows the detailed structure of our M-PSK receiver with USRP N200.

## III. USRP N200 AND SOLAR POWER CALCULATIONS

In this section we present the solar system power calculations associated with our USRP N200. From Table I, the circuit current consumption is 1.3 A and the RF current consumption is 2.3 A, while the DC input voltage is 6v. Based on these parameters, we can estimate the watt demand

for our system, where the total Watts Per Hour (DC) ($W_T$) is given by (1):

$$W_T = I_{DC} \times V_{DC} \quad (Watt) \quad (1)$$

The total daily usage Watt-hour per day ($DW_T$) is given by (2):

$$DW_T = W_T \times \#Hrs \ (watt-Hrs/d) \quad (2)$$

We assume battery depth of discharge, also known as DOD to be 20% in most deep cycle in order to get the best value for money. General rule of thumb: the less your deep cycle battery is discharged before being properly recharged again, the longer it will last. Table II presents the Battery State of Charge (SoC) for different types of batteries given voltage levels. Measuring SoC by voltage is simple, but it can be inaccurate because cell materials and temperature affect the voltage. The most blatant error of the voltage-based SoC occurs when disturbing a battery with a charge or discharge, which is the case in our real time system. For that, we use coulomb counting to estimate SoC by measuring the in-and-out-flowing current. Ampere-second is used for both charge and discharge. We compute SoC using equation (3), where $C_{nom}$ is the battery capacity in Ampere-hour, $i_{bat}(t)$ is the current drawn from the battery with negative sign because the battery is being depleted and $SOC(0)$ is the initial battery state of charge. Table III presents the complete solar power calculations [14] [15] [16] [17].

$$SOC = \frac{1}{C_{nom}} \int_0^t i_{bat}(t)dt + SOC(0) \quad (3)$$

TABLE II.  STATE OF CHARGE FOR DIFFERENT TYPES OF BATTERIES GIVEN VOLTAGE LEVELS

| State of Charge (SoC) | Sealed or Flooded Lead Acid (v) | Gel battery (v) | AGM battery (v) |
|---|---|---|---|
| 100% | 12.7+ | 12.85+ | 12.8+ |
| 75% | 12.4 | 12.65 | 12.6 |
| 50% | 12.3 | 12.35 | 12.3 |
| 25% | 12.0 | 12.0 | 12.0 |
| 0% | 11.8 | 11.8 | 11.8 |

## IV. EWS-SDR ALGORITHM

In this section we evaluate our system model described in section II, were we use M-PSK modulation techniques for proof of concept, namely, QPSK, 8-PSK, 16-PSK, 32-PSK and 64-PSK. An adaptive energy aware algorithm (EWS-SDR) has been proposed and tested over our system model.

Simulating the M-PSK modulation techniques over an AWGN channel for various SNRs to obtain the basic BER vs. SNR performance graphs as shown in Fig 4. Using our BER threshold of $10^{-4}$ we can divide the range of SNR in dB for which each of the modulation order is able to achieve our targeted BER. The shadowed areas in Fig.4 represent the required range of SNR for each modulation technique. Table IV summarizes the SNR ranges that will be used in our EWS-SRD algorithm.

TABLE III.  SOLAR SYSTEM CALCULATIONS WITH USRP N200

| | Calculations | | |
|---|---|---|---|
| | item | Value | Unit |
| Estimated Watts demand | Total Watts Per Hour (DC) | 13.8 | Watts |
| | Hours per day | 24 | Hrs/d |
| Watt-Hours per day | Total daily usage Watts x Hours | 331.2 | Watt-Hrs/d |
| Amp-hour | Total watts Daily requirements | 331.2 | Watt-Hrs/d |
| | Corrected for battery losses Assumes static average loss | 337.824 | Watt-Hrs/d |
| | System voltage DC voltage only | 6 | Volts |
| | Amp-hours per day Watts divided by Volts | 56.304 | Amp-Hrs/d |
| Battery bank calculation | # of days backup power required Average 24 hour periods | 1 | days |
| | Amp-hour storage required raw capacity | 56.304 | Amp-Hrs |
| | Depth of discharge | Assumes 20% | |
| | Required amp backup Prevents excessive discharge | 112.608 | Amp-Hrs |
| | Battery Amp Rating (20hr) Battery Capacity in Amps | 0.2 | fraction |
| | Batteries wired in series Relates to system voltage | 0.5 | |
| Solar Panel Array calculation | Sun hours per day (Direct only) | 7 | Hrs |
| | Worst-weather multiplier* 1.55 default | 1.55 | Fraction |
| | Total sun hours per day Assumes average sun | 4.516 | Amp-Hrs |
| | Select panel size (Watt rating) Watt hour rating | 100 | Watts |
| | Nominal Panel Voltage Approximate Solar output | 16 | Volts |
| | Amps required from solar panels Total daily consumption | 56 | Amps |
| | Peak amperage of solar panel Watts divided by Volts | 6.25 | Amps |
| | Number of solar panels in parallel Raw Number | 1.995 | |
| | Number of panels in series (12 V) it is 1 for 12v, 2 for 24v, etc | 1 | |
| | Rounded number of solar panels | 1 | |

Figure 5 describes the flow chart of the EWS-SDR algorithm. The proposed algorithm starts by calculating the SoC to check if the battery's charging percentage is higher than its deep depth, otherwise it will wait until the battery charges. If SoC is greater than 20%, then the system will follow the user preferences by choosing either to work in the CE mode or the

QE mode. In the later mode the user cares about the quality of the transmission more than staying connected for longer periods of time, mainly when the required Data Transmission Time (DTT) is short in duration, e.g. streaming short video. In this mode, the system will do its best to achieve the highest transmission quality. However, in QE mode the system will also take into consideration the available power and it will estimate the Available Transmission Time for each Modulation order (ATT_$Mod_i$). Based on ATT_$Mod_i$, the system will pick the highest modulation order ($Mod_i$) that has ATT_Mod higher than DTT. On contrary, in the CE mode, the user main concern is to stay connected as first priority and quality comes as an addition, e.g. data from security cameras; to have a bad quality video for longer time when the SoC is low is better than having a high-quality interrupted video. In the CE mode, the system will set the modulation order based on the available energy from the solar system, as much there is energy as higher order modulation is used and vice versa.

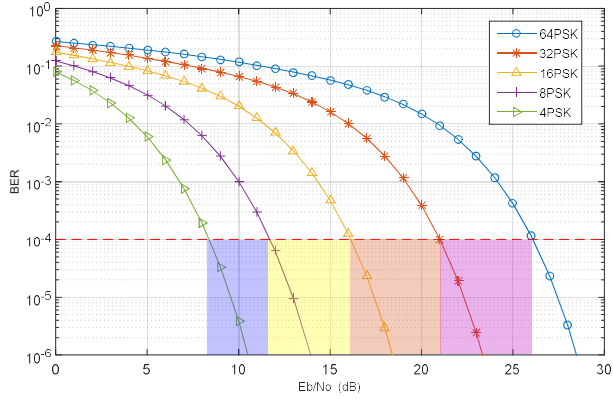

Fig. 4. SNR vs. BER for several MPSK modulations over AWGN channel

TABLE IV. MODULATION TYPE VS. REQUIRED SNR RANGE FOR M-PSK, WITH BER THRESHOLD OF $10^{-4}$

| Modulation | SNR Range (dB) |
|---|---|
| QPSK | 8 - 12 |
| 8-PSK | 12 - 16 |
| 16-PSK | 16 - 22 |
| 32-PSK | 22 - 27 |
| 64-PSK | > 27 |

## V. RESULTS

In this section we evaluate the system setup. From Fig 6, it is noticeable that the current consumption increases as the RF transmission power increases. By comparing the four levels of transmitter system gains, i.e. 0, 10, 20 and 30 dB, one can notice that there is a significant jump in the current consumption when the system gain is changed form 0 dB to 30 dB,

Figure 7, shows the current consumption for different system gain configurations at maximum power transmission, the figure shows that for a gain less than 15 dB the current consumption is almost linear with gain increment, on the other hand, a sharp increase in the current consumption is noticed when the system gain is set to 20 dB or higher, which confirms the results in Fig.6. By investigating Fig.6, we notice that when the transmission level is zero, there will be some current consumption due to power consumption of the device electronics.

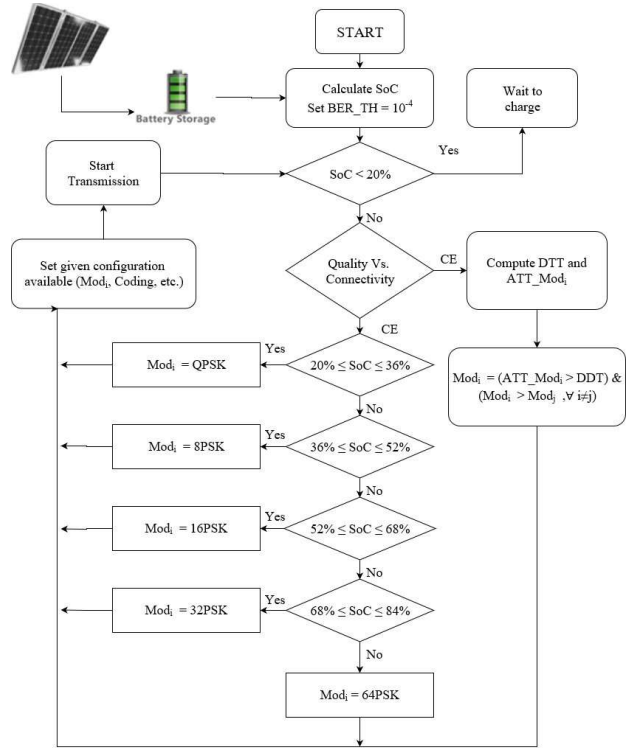

Fig. 5. Flow graph for the EWS-SDR algorithm

In Fig. 8, the SoC and the battery voltage characteristic for an initially fully charged battery is shown. Taking in account the current consumption shown in Fig.6, it is noticeable that the SoC and the battery voltage will decrease as the current consumption increases. Moreover, based on different system gain levels the SoC value varies widely, for example when the system gain increases form 0 dB to 30 dB, the SoC decreases by 3.5 %.

## VI. CONCLUSION

In this paper, an energy aware wireless system based on software defined radio is presented. Moreover, an adaptive modulation algorithm (EWS-SDR) is proposed. In the EWS-SDR algorithm, the system takes into consideration the available solar energy by checking the solar battery's state of charge before deciding on which modulation order to use. In addition, it allows the user to decide either to work in a connectivity efficient mode or in a quality efficient mode. Our results show how the current consumption will increase significantly when increasing the transmission power and the transmitter system gain. Moreover, this work offers graphs to characterize the energy consumption behavior of the USRP N200 software defined radio.

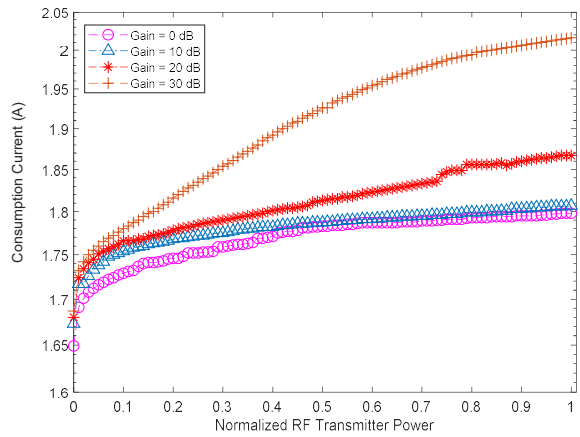

Fig. **6**. The current consumption vs. normalized RF transmission power for different levels of transmitter system gains.

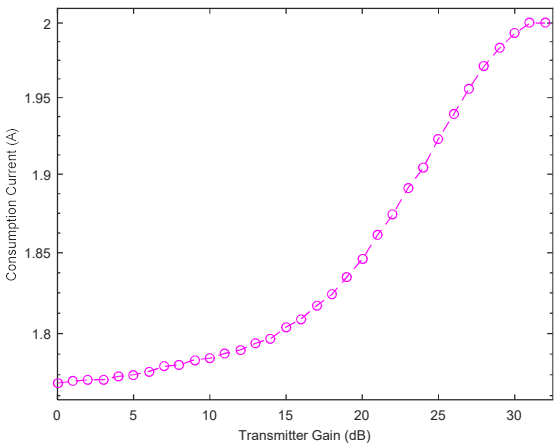

Fig. **7**. The current consumption vs. different levels of transmitter system gains while using the highest level of power.

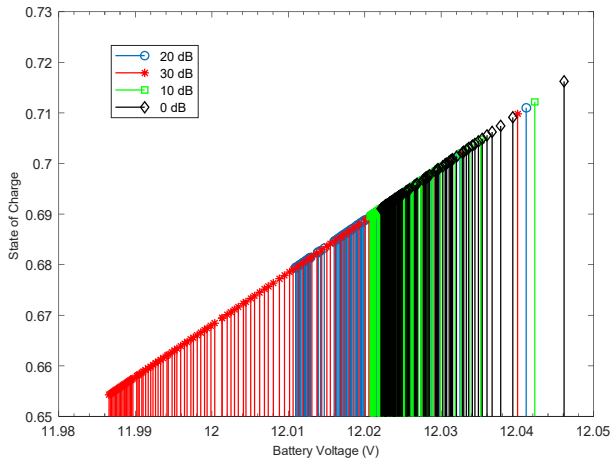

Fig. **8**. Battery state of charge for different USRP N200 system gains.